\documentclass[runningheads]{llncs}
\usepackage[T1]{fontenc}
\usepackage{graphicx}
\usepackage{xcolor}
\usepackage{amssymb}
\usepackage{amsmath}
\usepackage{braket}
\usepackage[acronym]{glossaries}
\makeglossaries

\newacronym{qubo}{QUBO}{quadratic unconstrained binary optimization}
\newacronym{qaoa}{QAOA}{quantum approximate optimization algorithm}
\newacronym{cvar}{CVaR}{conditional value at risk}
\newacronym{cvarqaoa}{CVaR-QAOA}{conditional value-at-risk quantum approximate optimization algorithm}
\newacronym{qml}{QML}{quantum machine learning}
\newacronym{cobyla}{COBYLA}{Constrained Optimization BY Linear Approximations}

\newacronym{rq1}{RQ1}{Research question 1}
\newacronym{rq2}{RQ2}{Research question 2}
\newacronym{rq3}{RQ3}{Research question 3}

\begin{document}
\title{Hybrid Quantum-Classical Optimization for Emergency-Priority Traffic Signal Control}
\author{%
Silvie Ill\'{e}sov\'{a}\inst{4}\thanks{Corresponding author.} \and
Vojt\v{e}ch Nov\'{a}k\inst{1,2,3}}
\authorrunning{S. Ill\'{e}sov\'{a} et al.}
\institute{%
Department of Computer Science, Faculty of Electrical Engineering and Computer Science,
VSB -- Technical University of Ostrava, Ostrava, Czech Republic
\and
IT4Innovations National Supercomputing Center, VSB -- Technical University of Ostrava,
708 00 Ostrava, Czech Republic
\and
Department of Informatics and Statistics, Marine Research Institute, Klaipeda University,
Klaipeda, Lithuania
\and
Gran Sasso Science Institute, L'Aquila, Italy\\
\email{silvie.illesova@gssi.it}
}
\titlerunning{Quantum Optimization for Emergency-Priority Traffic Control}
\maketitle              % typeset the header of the contribution
\begin{abstract}
Emergency-priority traffic signal control requires reducing emergency-vehicle delay while limiting disruption to ordinary traffic. This work investigates a finite-horizon QUBO formulation and compares Exact QUBO optimization, QAOA, CVaR-QAOA, a quantum machine-learning policy, fixed-time control, and max-pressure control in a three-intersection corridor. The study evaluates three traffic-demand regimes, examines the effect of the emergency-priority weight, and compares quantum solutions with the exact QUBO optimum. The optimization-based controllers achieved reliable emergency progression across all tested demand regimes, while max-pressure produced lower ordinary-traffic delay. Increasing emergency priority reduced emergency travel time but increased congestion, with diminishing improvements at larger weights. CVaR-QAOA reproduced the exact QUBO solution more accurately than standard QAOA, although the resulting differences in closed-loop traffic performance were comparatively small.

\end{abstract}
\section{Introduction}

Urban traffic-signal control is fundamentally a multi-objective problem in which reducing congestion must be balanced against additional operational priorities \cite{hamilton2013evolution,zhao2011computational,papageorgiou2003review}. Emergency vehicles introduce a particularly demanding case because signal decisions should reduce their delay without causing excessive disruption to ordinary traffic \cite{khan2018evp,yu2022state,karmakar2020smart}. Conventional fixed-time strategies cannot respond to such events, while adaptive approaches such as queue-responsive and max-pressure control react to observed traffic conditions but do not necessarily incorporate explicit emergency-priority objectives \cite{wang2018review,favilla1993fuzzy,pandit2013adaptive}.

Optimization-based traffic control provides a natural way to encode this trade-off by combining queue conditions, coordination requirements, switching penalties, and emergency priority within a common objective \cite{jin2017stochastic,mao2021boosted,xu2019optimizing}. In particular, \gls{qubo} \cite{punnen2022quadratic,lewis2017quadratic} formulations are attractive because discrete signal decisions can be represented directly as binary variables and solved using either classical or quantum optimization methods \cite{neukart2017traffic,singh2021quantum}. Variational quantum algorithms such as \gls{qaoa} \cite{zhou2020quantum} have therefore been investigated as candidate approaches for combinatorial optimization, while risk-sensitive variants based on conditional value at risk can concentrate optimization on lower-energy regions of the sampled distribution \cite{grange2023introduction,montanez2025toward,montanez2025transfer}. Quantum machine-learning approaches provide a complementary possibility in which a parameterized quantum model learns a control policy from optimization-generated examples \cite{navarro2022traffic,fadlullah2017state,khan2025smart,almukhalfi2024traffic}. More broadly, variational quantum algorithms have been developed and evaluated for a variety of problems, including ground- and excited-state electronic structure and near-degenerate quantum systems \cite{beseda2024state,illesova2025transformation,PhysRevA.111.022437}.

However, evaluating such methods for traffic control requires more than comparing optimization-objective values. In a receding-horizon controller, only the first decision of an optimized schedule is executed before the problem is rebuilt, and safety constraints can further modify the requested phase. Consequently, a more accurate solution of the underlying \gls{qubo} does not necessarily translate directly into improved closed-loop traffic performance. Moreover, emergency-priority control introduces an explicit trade-off between emergency progression and ordinary congestion that must be evaluated across different demand regimes and priority strengths. This work addresses these issues through a controlled comparison of classical baselines, Exact \gls{qubo} optimization, \gls{qaoa}, \gls{cvarqaoa} \cite{barkoutsos2020improving,yu2025improved,ruiz2026binary}, and a \gls{qml} policy in a coordinated three-intersection corridor.

\subsection{Research Questions}

This study addresses three research questions. \textbf{\glsentryshort{rq1}:} \emph{How effectively do the \gls{qubo}-based, quantum-optimization, and quantum-learning controllers reduce emergency-vehicle delay while maintaining ordinary traffic performance compared with fixed-time and max-pressure control?} \textbf{\glsentryshort{rq2}:} \emph{How does the emergency-priority weight affect the trade-off between emergency-vehicle progression and ordinary traffic congestion?} \textbf{\glsentryshort{rq3}:} \emph{To what extent do \gls{qaoa} and \gls{cvarqaoa} approximate the exact finite-horizon \gls{qubo} solution, and how strongly do \gls{qubo} solution errors influence traffic-control performance?}

\subsection{Contributions}

The main contributions of this work are a finite-horizon \gls{qubo} formulation for coordinated emergency-priority traffic control; a comparison of Exact \gls{qubo}, \gls{qaoa}, \gls{cvarqaoa}, and \gls{qml} with fixed-time and max-pressure baselines; evaluation under balanced, main-road-dominant, and side-street-dominant traffic demand; analysis of the emergency-priority coefficient $\lambda_{\mathrm{em}}$ and step-level comparison of \gls{qaoa} and \gls{cvarqaoa} with exact \gls{qubo} solutions.

% ============================================================
\section{Background and Related Work}
\label{sec:background}
% ============================================================
Traffic signal control has evolved from fixed-time scheduling toward adaptive and optimization-based strategies that respond to changing traffic conditions. In the context of emergency-priority control, these approaches must balance rapid emergency-vehicle progression with the need to maintain acceptable performance for ordinary traffic. This section reviews the main classical and quantum optimization concepts relevant to the control methods evaluated in this work.

\subsection{Adaptive Traffic Signal Control}

Traffic-signal control ranges from fixed-time schedules \cite{serafini1989mathematical,thunig2019optimization} to adaptive queue-responsive methods \cite{albdairi2026queue,cai2009performance}. Fixed-time control is simple and reproducible but does not react to changing demand or emergency vehicles. Adaptive controllers \cite{samadi2012performance,gartner2002optimized} instead use the current traffic state to choose phases, while emergency-priority control \cite{devika2024next,rosayyan2023optimal} additionally seeks to reduce emergency waiting and travel time. This introduces a trade-off between emergency progression and ordinary traffic congestion.

This work compares fixed-time and queue-responsive classical baselines with exact \gls{qubo} optimization, \gls{qaoa}, \gls{cvarqaoa}, and a quantum machine-learning policy. All controllers use the same safety supervisor.

\subsection{\glsentryshort{qubo}-Based Optimization}

A quadratic unconstrained binary optimization problem \cite{lewis2017quadratic,punnen2022quadratic} uses binary variables $x_j\in\{0,1\}$ and minimizes
\begin{equation}
E(\mathbf{x})=
\sum_{j=0}^{n-1} l_j x_j
+
\sum_{0\leq j<r<n}Q_{jr}x_jx_r,
\label{eq:general_qubo}
\end{equation}
where $\mathbf{x}=(x_0,\ldots,x_{n-1})$ is the binary decision vector, $n$ is the number of variables, $l_j$ is the linear coefficient of $x_j$, and $Q_{jr}$ is the quadratic coupling between $x_j$ and $x_r$.

In this work, the binary variables represent traffic-light decisions over a finite prediction horizon.

\subsection{Quantum Approximate Optimization Algorithm}

The \gls{qubo} is mapped to Ising variables through
\begin{equation}
x_j=\frac{1-z_j}{2},
\qquad
z_j\in\{-1,+1\},
\end{equation}
where $z_j$ is the spin variable corresponding to binary variable $x_j$.

The \gls{qaoa} cost Hamiltonian \cite{choi2019tutorial} is
\begin{equation}
H_C=
\sum_{j=0}^{n-1}h_j Z_j
+
\sum_{0\leq j<r<n}J_{jr}Z_jZ_r,
\end{equation}
where $h_j$ and $J_{jr}$ are the Ising linear and interaction coefficients, and $Z_j$ is the Pauli-$Z$ operator acting on qubit $j$. The mixer Hamiltonian is
\begin{equation}
H_M=\sum_{j=0}^{n-1}X_j,
\end{equation}
where $X_j$ is the Pauli-$X$ operator on qubit $j$.

For \gls{qaoa} depth $p$, the variational state is
\begin{equation}
|\psi(\boldsymbol{\gamma},\boldsymbol{\beta})\rangle
=
\prod_{\ell=1}^{p}
e^{-i\beta_{\ell}H_M}
e^{-i\gamma_{\ell}H_C}
|+\rangle^{\otimes n},
\end{equation}
where $\boldsymbol{\gamma}=(\gamma_1,\ldots,\gamma_p)$ and $\boldsymbol{\beta}=(\beta_1,\ldots,\beta_p)$ are trainable \gls{qaoa} parameters.

Standard \gls{qaoa} minimizes the expected \gls{qubo} energy
\begin{equation}
F_{\mathrm{qaoa}}(\boldsymbol{\theta})
=
\sum_{\mathbf{x}}
P_{\boldsymbol{\theta}}(\mathbf{x})E(\mathbf{x}),
\end{equation}
where $\boldsymbol{\theta}=(\boldsymbol{\gamma},\boldsymbol{\beta})$ and $P_{\boldsymbol{\theta}}(\mathbf{x})$ is the probability assigned to bitstring $\mathbf{x}$.

Variational quantum algorithms rely on a hybrid optimization loop in which
the performance of the quantum model can depend strongly on the choice of
classical optimizer, parameterization, and the presence of sampling or hardware noise \cite{novak2025optimization,saoovqebenchmarking,novak2025reliable,bezdvek2025classical}. Systematic comparisons of optimization strategies have therefore become an important component of evaluating variational quantum methods \cite{novak2026longitudinal}. \gls{qaoa} has also been investigated outside its original combinatorial-optimization setting, including applications such as software-engineering test-case selection \cite{trovato2025preliminary}.

\subsection{\gls{cvarqaoa}}

\gls{cvarqaoa} \cite{barkoutsos2020improving} uses the same circuit as standard \gls{qaoa} but optimizes only the lowest-energy fraction of the output probability distribution.

Let the bitstrings be ordered so that
\begin{equation}
E(\mathbf{x}^{(1)})
\leq
E(\mathbf{x}^{(2)})
\leq
\cdots
\leq
E(\mathbf{x}^{(2^n)}),
\end{equation}
and let $p_r=P_{\boldsymbol{\theta}}(\mathbf{x}^{(r)})$. For CVaR fraction $\alpha_{\mathrm{CVaR}}\in(0,1]$, the probability mass taken from state $r$ is
\begin{equation}
w_r=
\min
\left\{
p_r,
\max
\left[
0,
\alpha_{\mathrm{CVaR}}
-
\sum_{s=1}^{r-1}w_s
\right]
\right\}.
\end{equation}
The CVaR objective is
\begin{equation}
F_{\mathrm{CVaR}}(\boldsymbol{\theta})
=
\frac{
\sum_r w_rE(\mathbf{x}^{(r)})
}{
\max\left\{\sum_r w_r,10^{-15}\right\}
}.
\end{equation}
The default value used in the experiments is $\alpha_{\mathrm{CVaR}}=0.20$.

\subsection{Quantum Machine Learning for Control}

Parameterized quantum models have been investigated across a range of
classification and hybrid-learning settings, with recent studies emphasizing the importance of circuit architecture, feature representation, trainability, and the interaction between classical and quantum components \cite{illesova2025qmetric,illesova2026importance,illesova2025classical,illesova2025complementarity}. Quantum neural-network approaches have also been explored in applied prediction problems, including clinical classification tasks \cite{novak2026quantum}.

In this work, the \gls{qml} controller is trained offline to imitate the first action chosen by the exact \gls{qubo} controller. It uses one qubit per intersection and receives queue, phase, phase-age, and emergency-proximity features. Its trainable parameters remain fixed during traffic simulation.

% ============================================================
\section{Traffic-Control Model}
\label{sec:traffic_model}
% ============================================================
The traffic-control problem is formulated on a signalized corridor in which vehicle queues, signal phases, and emergency-vehicle progression evolve over discrete simulation steps. The model provides the common traffic state and decision representation used by all controllers, allowing their behavior to be compared under identical traffic and safety conditions.
\subsection{Corridor and Traffic State}

The corridor contains $N$ signalized intersections. At simulation step $k$, the main-road and side-street queue lengths at intersection $i$ are
\begin{equation}
q_i^M(k),
\qquad
q_i^S(k),
\end{equation}
where $i\in\{0,\ldots,N-1\}$, superscripts $M$ and $S$ denote the main road and side street, respectively, and $k$ denotes the simulation step.

An emergency vehicle travels along the main corridor. Its movement and waiting are simulated explicitly, while queue lengths are updated by the traffic-flow model.

\subsection{Signal Decisions}

Each controller returns a desired binary phase
\begin{equation}
d_i(k)=
\begin{cases}
1, & \text{main-road green},\\
0, & \text{side-street green},
\end{cases}
\end{equation}
where $d_i(k)$ is the requested phase at intersection $i$ and step $k$.

The safety supervisor converts this request into the actuated phase
\begin{equation}
u_i(k)\in\{-1,0,1\},
\end{equation}
where $u_i(k)=-1$ denotes all-red, $u_i(k)=0$ side-street green, and $u_i(k)=1$ main-road green.

\subsection{Safety Supervisor}

For each intersection, the supervisor stores the current phase, its age, an all-red countdown, and an optional pending phase. A phase change is allowed only when
\begin{equation}
A_i(k)\geq G_{\min},
\end{equation}
where $A_i(k)$ is the age of the current green phase and $G_{\min}$ is the minimum-green duration.

The experiments use
\begin{equation}
G_{\min}=4,
\qquad
R=1,
\end{equation}
where $R$ is the number of all-red transition steps.

% ============================================================
\section{Control Methods}
\label{sec:control_methods}
% ============================================================
The evaluated control strategies range from simple classical baselines to optimization-based and quantum approaches. Each controller receives the same traffic state and produces desired signal-phase decisions, which are subsequently processed by the common safety supervisor described in Section~\ref{sec:traffic_model}. This section defines the individual control methods and their implementation within the experimental framework.
\subsection{Fixed-Time Controller}

The fixed-time controller alternates between main-road and side-street green in blocks of length $G_{\min}$. The block index is
\begin{equation}
b(k)=
\left\lfloor
\frac{k}{\max\{1,G_{\min}\}}
\right\rfloor,
\end{equation}
and the desired phase is
\begin{equation}
d_i(k)=
\begin{cases}
1, & b(k)\bmod 2=0,\\
0, & b(k)\bmod 2=1.
\end{cases}
\end{equation}
Here, $b(k)$ is the current timing block.

\subsection{Simplified Max-Pressure Controller}

The implemented max-pressure baseline compares the local queues:
\begin{equation}
d_i(k)=
\begin{cases}
1, & q_i^M(k)\geq q_i^S(k),\\
0, & q_i^M(k)<q_i^S(k).
\end{cases}
\end{equation}
Ties are resolved in favor of the main road. The controller does not use downstream pressure, turning ratios, prediction horizons, or emergency information.

\subsection{Finite-Horizon \glsentryshort{qubo} Controller}

\subsubsection{Decision Variables}

For intersection $i$ and horizon position $t$, the decision variable is
\begin{equation}
x_{i,t}\in\{0,1\},
\end{equation}
where $x_{i,t}=1$ requests main-road green and $x_{i,t}=0$ requests side-street green.

The pair $(i,t)$ is flattened as
\begin{equation}
j(i,t)=iT+t,
\end{equation}
where $T$ is the optimization horizon length. Therefore,
\begin{equation}
n=NT
\end{equation}
is the total number of \gls{qubo} variables.

\subsubsection{Queue-Pressure Term}

The predicted queues are initialized as
\begin{equation}
\widehat q_i^M(0)=q_i^M(k),
\qquad
\widehat q_i^S(0)=q_i^S(k),
\end{equation}
where $\widehat q_i^M(t)$ and $\widehat q_i^S(t)$ are predicted queues at horizon position $t$.

The queue contribution is
\begin{equation}
c_{i,t}^{\mathrm{queue}}
=
\beta\widehat q_i^S(t)
-
\alpha\widehat q_i^M(t),
\end{equation}
where $\alpha$ and $\beta$ weight main-road and side-street pressure.

The predicted queues are updated using expected arrivals,
\begin{equation}
\widehat q_i^M(t+1)
=
\widehat q_i^M(t)
+
p_M^{\mathrm{exp}}b_M^{\mathrm{exp}}\eta_i,
\end{equation}
\begin{equation}
\widehat q_i^S(t+1)
=
\widehat q_i^S(t)
+
p_S^{\mathrm{exp}}b_S^{\mathrm{exp}},
\end{equation}
where $p_M^{\mathrm{exp}}$ and $p_S^{\mathrm{exp}}$ are expected arrival probabilities, $b_M^{\mathrm{exp}}$ and $b_S^{\mathrm{exp}}$ are expected arrival batch sizes, and
\begin{equation}
\eta_i=
\begin{cases}
1, & i=0,\\
0.5, & i>0
\end{cases}
\end{equation}
scales expected downstream main-road arrivals.

\subsubsection{Signal-Age Term}

The phase-age adjustment is
\begin{equation}
a_i=
\lambda_{\mathrm{age}}
\max\{0,A_i(k)-G_{\min}+1\},
\end{equation}
where $\lambda_{\mathrm{age}}$ controls the strength of the phase-age penalty.

Its \gls{qubo} contribution is
\begin{equation}
c_{i,t}^{\mathrm{age}}
=
\begin{cases}
+a_i, & \text{current main-road green},\\
-a_i, & \text{current side-street green}.
\end{cases}
\end{equation}

\subsubsection{Emergency-Priority Term}

Let $c_{\mathrm{em}}(k)$ be the emergency-vehicle cell and $c_i^{\mathrm{stop}}$ the stop-line cell of intersection $i$. The distance is
\begin{equation}
d_i^{\mathrm{em}}(k)
=
c_i^{\mathrm{stop}}
-
c_{\mathrm{em}}(k).
\end{equation}
The target horizon position is
\begin{equation}
\tau_i(k)
=
\max\{0,d_i^{\mathrm{em}}(k)\},
\end{equation}
where $\tau_i(k)$ approximates when the vehicle reaches intersection $i$.

The emergency reward is
\begin{equation}
r_{i,t}^{\mathrm{em}}
=
\begin{cases}
\dfrac{\lambda_{\mathrm{em}}}
{1+|t-\tau_i(k)|},
&
|t-\tau_i(k)|\leq m_{\mathrm{em}},
\\[6pt]
0,
&
\text{otherwise},
\end{cases}
\end{equation}
where $\lambda_{\mathrm{em}}$ is the emergency-priority weight and $m_{\mathrm{em}}$ is the emergency look-ahead margin.

\subsubsection{Switching Penalty}

Changes between consecutive horizon decisions are penalized by
\begin{equation}
E_{\mathrm{sw}}(\mathbf{x})
=
\lambda_{\mathrm{sw}}
\sum_{i=0}^{N-1}
\sum_{t=1}^{T-1}
(x_{i,t}-x_{i,t-1})^2,
\end{equation}
where $\lambda_{\mathrm{sw}}$ is the switching-penalty weight.

\subsubsection{Coordination Reward}

Neighboring main-road greens are rewarded using
\begin{equation}
w_i=
\lambda_{\mathrm{coord}}
\left(
1+\min\{\widehat q_i^M(T),s_M\}
\right),
\end{equation}
where $\lambda_{\mathrm{coord}}$ is the coordination weight and $s_M$ is the main-road service capacity.

The coordination contribution is
\begin{equation}
E_{\mathrm{coord}}(\mathbf{x})
=
-
\sum_{i=0}^{N-2}
w_i
\sum_{t=0}^{T-1}
x_{i,t}x_{i+1,t}.
\end{equation}

\subsubsection{Supervisor-Aware First Action}

If the signal cannot switch immediately, the first \gls{qubo} action is strongly biased toward the current phase using
\begin{equation}
M=1000,
\end{equation}
where $M$ is the large penalty magnitude.

The corresponding first-step coefficient is
\begin{equation}
c_{i,0}^{\mathrm{sup}}
=
\begin{cases}
-M, & \text{current main-road green},\\
+M, & \text{current side-street green}.
\end{cases}
\end{equation}

\subsubsection{Receding-Horizon Application}

Only the first optimized action is applied:
\begin{equation}
d_i(k)=x_{i,0}^{\star},
\end{equation}
where $x_{i,0}^{\star}$ is the optimized first-horizon decision. The remaining horizon decisions are discarded and the \gls{qubo} is rebuilt at the next simulation step.

\subsection{Difference Between Exact \glsentryshort{qubo}, \glsentryshort{qaoa}, and \glsentryshort{cvarqaoa}}

Exact \gls{qubo}, \gls{qaoa}, and \gls{cvarqaoa} optimize the same finite-horizon \gls{qubo} objective but differ in how they search the binary solution space. Exact \gls{qubo} exhaustively evaluates all candidate bitstrings and therefore returns a globally optimal solution of the formulated \gls{qubo}. Standard \gls{qaoa} instead optimizes a parameterized quantum state by minimizing the expected \gls{qubo} energy over the complete output probability distribution. \gls{cvarqaoa} uses the same quantum circuit and solution-selection procedure as standard \gls{qaoa}, but its variational objective is computed only from the lowest-energy fraction of the probability distribution. In the experiments, this fraction is $\alpha_{\mathrm{CVaR}}=0.20$. Thus, Exact \gls{qubo} provides the reference optimum, while \gls{qaoa} and \gls{cvarqaoa} provide approximate solutions to the same optimization problem using different variational objectives.

\subsection{Exact \glsentryshort{qubo} Solver}

The exact reference controller solves each finite-horizon \gls{qubo} by exhaustive enumeration of all binary assignments. For a problem with $n=NT$ binary variables, every bitstring $\mathbf{x}\in\{0,1\}^{n}$ is evaluated using the original \gls{qubo} energy, and the minimum-energy solution is selected:
\begin{equation}
\mathbf{x}^{\star}
=
\underset{\mathbf{x}\in\{0,1\}^{NT}}
{\operatorname{arg\,min}}
E(\mathbf{x}).
\end{equation}

In the implementation, binary schedules are enumerated as integer bit masks from $0$ to $2^{NT}-1$. For each mask, the energy is computed directly from the linear and quadratic \gls{qubo} coefficients. The minimum-energy mask is identified by exhaustive comparison and converted back to the binary decision vector $\mathbf{x}^{\star}$. Solutions whose energies agree with the minimum within a tolerance of $10^{-9}$ are treated as degenerate optima.

For the experimental values $N=3$ and $T=3$,
\begin{equation}
2^{NT}=2^9=512
\end{equation}
candidate schedules are evaluated at every control step. Because the controller operates in a receding-horizon manner, only the first decision for each intersection,
\begin{equation}
d_i(k)=x_{i,0}^{\star},
\end{equation}
is applied before the \gls{qubo} is rebuilt from the updated traffic state at the next simulation step.
% \subsection{Exact \glsentryshort{qubo} Solver}

% The exact reference controller evaluates all binary schedules and returns
% \begin{equation}
% \mathbf{x}^{\star}
% =
% \underset{\mathbf{x}\in\{0,1\}^{NT}}
% {\operatorname{arg\,min}}
% E(\mathbf{x}).
% \end{equation}
% For the experimental values $N=3$ and $T=3$,
% \begin{equation}
% 2^{NT}=2^9=512
% \end{equation}
% candidate schedules are evaluated.

\subsection{\glsentryshort{qaoa} Controller}

The \gls{qaoa} controller uses statevector simulation, circuit depth $p=2$, \gls{cobyla} optimization, five random restarts, and a warm start from the previous simulation step.

After optimization, the $K=64$ most probable bitstrings are retained and evaluated using the original \gls{qubo} energy. The selected state is
\begin{equation}
\mathbf{x}^{Q}
=
\underset{\mathbf{x}\in\mathcal{T}_K}
{\operatorname{arg\,min}}
E(\mathbf{x}),
\end{equation}
where $\mathcal{T}_K$ is the set of the $K$ most probable states.

The optimality gap is
\begin{equation}
\Delta E
=
E(\mathbf{x}^{Q})
-
E(\mathbf{x}^{\star}),
\end{equation}
where $\mathbf{x}^{Q}$ is the \gls{qaoa}-selected state and $\mathbf{x}^{\star}$ is the exact \gls{qubo} optimum.

\subsection{\glsentryshort{cvarqaoa} Controller}

\gls{cvarqaoa} uses the same \gls{qubo}, circuit, optimizer, restarts, warm start, and top-$K$ postselection as standard \gls{qaoa}. Only the variational objective differs.

The default CVaR fraction is
\begin{equation}
\alpha_{\mathrm{CVaR}}=0.20,
\end{equation}
so optimization focuses on the lowest-energy 20\% of the output probability mass.

\subsection{Quantum Machine-Learning Policy}

\subsubsection{Input Features}

For each intersection $i$, the \gls{qml} policy uses five features:
\begin{equation}
f_{i,1}
=
\tanh
\left(
\frac{q_i^M(k)-q_i^S(k)}{s_q}
\right),
\end{equation}
\begin{equation}
f_{i,2}
=
\tanh
\left(
\frac{q_i^M(k)+q_i^S(k)}{s_q}
\right),
\end{equation}
where $s_q$ is the queue-normalization scale.

The current phase is encoded as
\begin{equation}
f_{i,3}=
\begin{cases}
+1, & \text{main-road green},\\
-1, & \text{side-street green},
\end{cases}
\end{equation}
and the phase-age feature is
\begin{equation}
f_{i,4}
=
\tanh
\left(
\frac{A_i(k)}{\max\{1,G_{\min}\}}
\right).
\end{equation}
Emergency proximity is represented by
\begin{equation}
f_{i,5}
=
\exp
\left(
-
\frac{
|c_i^{\mathrm{stop}}-c_{\mathrm{em}}(k)|
}{
\delta_{\mathrm{em}}
}
\right),
\end{equation}
where $\delta_{\mathrm{em}}$ controls the spatial decay of the emergency feature.

\subsubsection{Policy Output}

The probability of requesting main-road green at intersection $i$ is
\begin{equation}
p_i^M
=
\sum_{\mathbf{z}:z_i=1}
|\langle\mathbf{z}|
\psi(\mathbf{f},\boldsymbol{\theta})
\rangle|^2,
\end{equation}
where $\mathbf{f}$ is the feature vector, $\boldsymbol{\theta}$ contains trainable circuit parameters, and $p_i^M$ is the marginal probability of qubit $i$ being in state $|1\rangle$.

The requested phase is
\begin{equation}
d_i(k)=
\begin{cases}
1, & p_i^M\geq0.5,\\
0, & p_i^M<0.5.
\end{cases}
\end{equation}

\subsubsection{Training Objective}

Exact \gls{qubo} decisions provide the labels
\begin{equation}
y_i=x_{i,0}^{\star},
\end{equation}
where $y_i$ is the target action for intersection $i$.

For $M_{\mathrm{tr}}$ training samples, the loss is
\begin{equation}
\mathcal{L}(\boldsymbol{\theta})
=
\frac{1}{M_{\mathrm{tr}}}
\sum_{m=1}^{M_{\mathrm{tr}}}
\left[
\frac{1}{N}
\sum_{i=0}^{N-1}
\left(
p_i^{(m)}-y_i^{(m)}
\right)^2
\right]
+
\lambda_{\mathrm{reg}}
\frac{1}{n_\theta}
\sum_{j=1}^{n_\theta}\theta_j^2,
\end{equation}
where $M_{\mathrm{tr}}$ is the number of training samples, $p_i^{(m)}$ is the predicted main-road probability for sample $m$, $y_i^{(m)}$ is the target label, $n_\theta$ is the number of trainable parameters, and $\lambda_{\mathrm{reg}}$ is the regularization coefficient.

The experiments use 32 synthetic training samples, two trainable layers, 50 \gls{cobyla} iterations, and $\lambda_{\mathrm{reg}}=10^{-3}$.
% ============================================================
\section{Experimental Design}
\label{sec:experimental_design}
% ============================================================
The experimental evaluation is designed to compare the considered controllers under a common traffic environment, parameterization, and set of stochastic traffic realizations. The experiments are organized around the three research questions, covering controller performance across traffic-demand regimes, the effect of emergency-priority weighting, and the approximation quality of the quantum optimization methods.
\subsection{General Configuration}
\label{subsec:general_configuration}

All experiments use a corridor with
\begin{equation}
N=3,
\qquad
T=3,
\qquad
NT=9,
\end{equation}
where $N$ is the number of signalized intersections, $T$ is the finite optimization horizon, and $NT$ is the number of binary \gls{qubo} variables.

The \glsentryshort{rq1} experiment uses 30 traffic-seed identifiers and 36 simulation steps. A longer-horizon robustness experiment is additionally performed for \glsentryshort{rq1} using 72 simulation steps and 10 traffic seeds. \glsentryshort{rq2} uses 30 seeds under balanced traffic demand and compares Exact \gls{qubo}, \gls{qaoa}, and \gls{cvarqaoa} while varying the emergency-priority weight. \glsentryshort{rq3} uses the step-level diagnostics recorded during the \gls{qaoa} and \gls{cvarqaoa} runs of \glsentryshort{rq1} and compares both quantum methods with the exact \gls{qubo} solution at every optimization step.

\subsection{Traffic-Demand Regimes}
\label{subsec:demand_regimes}

\glsentryshort{rq1} evaluates three traffic-demand regimes. Let $P_M$ denote the main-road arrival probability and $P_S$ the side-street arrival probability.
\begin{align}
\text{Balanced:}\qquad
&P_M=0.45,
&
P_S=0.45,
\\
\text{Main-road dominant:}\qquad
&P_M=0.60,
&
P_S=0.30,
\\
\text{Side-street dominant:}\qquad
&P_M=0.30,
&
P_S=0.60.
\end{align}
The same side-street arrival probability is used at all three intersections within a given demand regime. Arrival batches contain two vehicles for both the main road and side streets.

\subsection{Default \glsentryshort{qubo} Parameters}
\label{subsec:default_qubo_parameters}

Unless varied explicitly in \glsentryshort{rq2}, the \gls{qubo} parameters are
\begin{equation}
\alpha=1.0,
\qquad
\beta=0.75,
\qquad
\lambda_{\mathrm{sw}}=0.08,
\end{equation}
and
\begin{equation}
\lambda_{\mathrm{coord}}=1.8,
\qquad
\lambda_{\mathrm{age}}=0.20,
\qquad
\lambda_{\mathrm{em}}=35.
\end{equation}
Here, $\alpha$ and $\beta$ weight the main-road and side-street queue terms, respectively; $\lambda_{\mathrm{sw}}$ controls the switching penalty; $\lambda_{\mathrm{coord}}$ controls neighboring-intersection coordination; $\lambda_{\mathrm{age}}$ weights the signal-age term; and $\lambda_{\mathrm{em}}$ controls emergency-vehicle priority.

The safety supervisor uses
\begin{equation}
G_{\min}=4,
\qquad
R=1,
\end{equation}
where $G_{\min}$ is the minimum-green duration and $R$ is the all-red transition duration.

\subsection{Quantum-Optimization Parameters}
\label{subsec:quantum_parameters}

The \gls{qaoa} and \gls{cvarqaoa} experiments use
\begin{equation}
p=2,
\qquad
R_{\mathrm{opt}}=5,
\qquad
K=64,
\qquad
\alpha_{\mathrm{CVaR}}=0.20,
\end{equation}
where $p$ is the \gls{qaoa} circuit depth, $R_{\mathrm{opt}}$ is the number of random optimization restarts, $K$ is the number of highest-probability bitstrings retained for postselection, and $\alpha_{\mathrm{CVaR}}$ is the probability fraction used by the CVaR objective.

The classical parameter optimizer is \gls{cobyla} with a maximum of 60 iterations. From the second traffic-control step onward, the optimized parameters from the preceding step are also included as a warm-start candidate.

\subsection{\glsentryshort{rq1}: Controller Comparison}
\label{subsec:rq1_design}

\glsentryshort{rq1} compares six controllers: fixed-time control, max-pressure control, Exact \gls{qubo} optimization, \gls{qaoa}, \gls{cvarqaoa}, and the \gls{qml} policy. All six controllers are evaluated under balanced, main-road-dominant, and side-street-dominant traffic demand using the same set of 30 traffic-seed identifiers.

The recorded outcomes are emergency completion, emergency waiting time, emergency travel time, cumulative delay proxy, mean main-road queue length, mean side-street queue length, the number of completed signal switches, and the number of all-red intersection steps. The 72-step robustness experiment repeats the same controller comparison using 10 traffic seeds.

\subsection{\glsentryshort{rq2}: Emergency-Priority Weight}
\label{subsec:rq2_design}

\glsentryshort{rq2} evaluates the effect of the emergency-priority coefficient under balanced demand. The tested values are
\begin{equation}
\lambda_{\mathrm{em}}
\in
\{0,5,10,20,35,50,75\}.
\label{eq:rq2_lambda_values}
\end{equation}
For each value, Exact \gls{qubo}, \gls{qaoa}, and \gls{cvarqaoa} are evaluated using the same 30 traffic-seed identifiers. The recorded outcomes are emergency waiting time, emergency travel time, cumulative delay proxy, and mean side-street queue length.

\subsection{\glsentryshort{rq3}: Quantum Approximation Quality}
\label{subsec:rq3_design}

\glsentryshort{rq3} compares \gls{qaoa} and \gls{cvarqaoa} with the exact \gls{qubo} solution using the step-level diagnostics recorded during the \glsentryshort{rq1} simulations.

For optimization step $k$, the \gls{qubo} optimality gap is
\begin{equation}
\Delta E(k)
=
E_k\!\left(\mathbf{x}^{Q}(k)\right)
-
E_k\!\left(\mathbf{x}^{\star}(k)\right),
\end{equation}
where $\mathbf{x}^{Q}(k)$ is the bitstring selected by \gls{qaoa} or \gls{cvarqaoa} and $\mathbf{x}^{\star}(k)$ is an exact \gls{qubo} optimum.

An exact-solution hit is recorded when
\begin{equation}
|\Delta E(k)|\leq 10^{-9}.
\end{equation}
The first-action agreement is
\begin{equation}
A_{\mathrm{first}}(k)
=
\frac{1}{N}
\sum_{i=0}^{N-1}
\mathbb{I}
\left[
x^{Q}_{i,0}(k)
=
x^{\star}_{i,0}(k)
\right],
\end{equation}
where $\mathbb{I}[\cdot]$ is the indicator function.

The \glsentryshort{rq3} analysis evaluates the mean \gls{qubo} optimality gap, exact-solution hit rate, mean probability assigned to globally optimal states, mean probability of the selected state, first-action agreement, emergency travel time, and cumulative delay proxy.

\subsection{Statistical Analysis}
\label{subsec:statistical_analysis}

For each reported continuous metric, results are summarized using the sample mean across available runs. Error bars represent two-sided 95\% confidence intervals for the mean, calculated as
\begin{equation}
\bar{x}
\pm
t_{0.975,n-1}
\frac{s}{\sqrt{n}},
\end{equation}
where $\bar{x}$ is the sample mean, $s$ is the sample standard deviation computed with $n-1$ degrees of freedom, $n$ is the number of available observations, and $t_{0.975,n-1}$ is the corresponding Student's $t$ critical value. Missing values are excluded from the calculation; consequently, confidence intervals for emergency travel time are based only on runs in which the emergency vehicle completed the corridor.

\section{Results}
The results are organized according to the three research questions. We first compare the overall traffic-control performance of all evaluated controllers, then examine the effect of the emergency-priority coefficient, and finally analyze how closely the quantum optimization methods reproduce the exact \gls{qubo} solution.
\subsection{\glsentryshort{rq1}: Comparison of Control Methods}
\glsentryshort{rq1} compares the six control strategies across the three traffic-demand regimes with respect to both emergency-vehicle progression and ordinary-traffic performance. The analysis first considers emergency completion and travel time, then examines cumulative delay and queue behavior, before discussing the resulting emergency--congestion trade-off, the performance of the \gls{qml} policy, and the longer-horizon robustness experiment.

\subsubsection{Emergency-Vehicle Performance}

\begin{figure}[t]
    \centering
    \includegraphics[width=\linewidth]{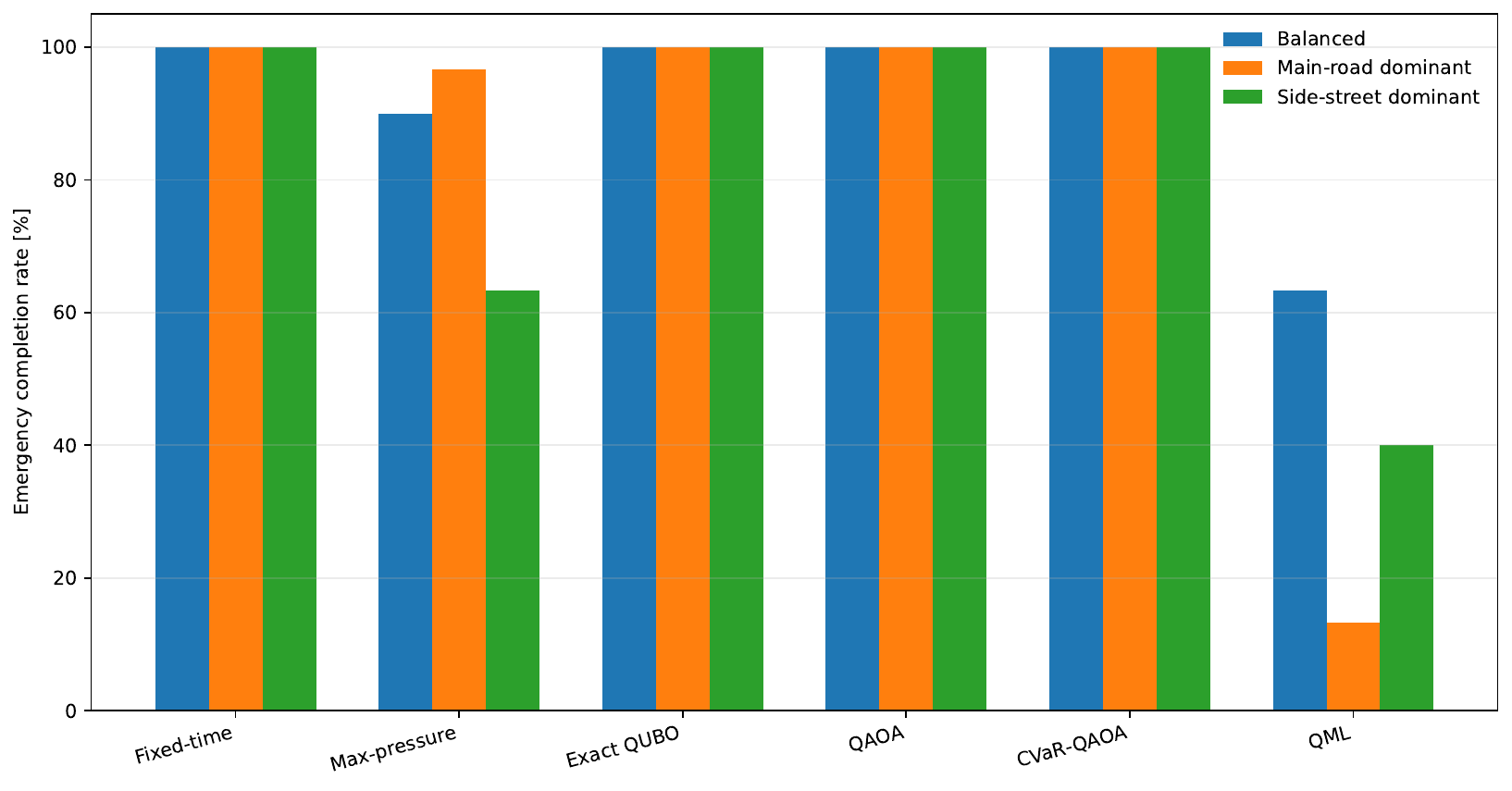}
    \caption{Emergency-vehicle completion rate for each controller under the three traffic-demand regimes.}
    \label{fig:rq1_emergency_completion}
\end{figure}

The optimization-based controllers provided the most reliable emergency
progression across the three demand regimes. Exact \gls{qubo}, \gls{qaoa}, and
\gls{cvarqaoa} achieved a 100\% emergency-completion rate in all tested
conditions, as shown in Fig.~\ref{fig:rq1_emergency_completion}. Fixed-time
control also completed all emergency trips, whereas max-pressure failed to
complete the trip within the 36-step horizon in several runs, particularly
under side-street-dominant demand. The \gls{qml} policy was substantially less
reliable, with completion rates of 63.3\%, 13.3\%, and 40.0\% under
balanced, main-road-dominant, and side-street-dominant demand,
respectively.

\begin{figure}[t]
    \centering
    \includegraphics[width=\linewidth]{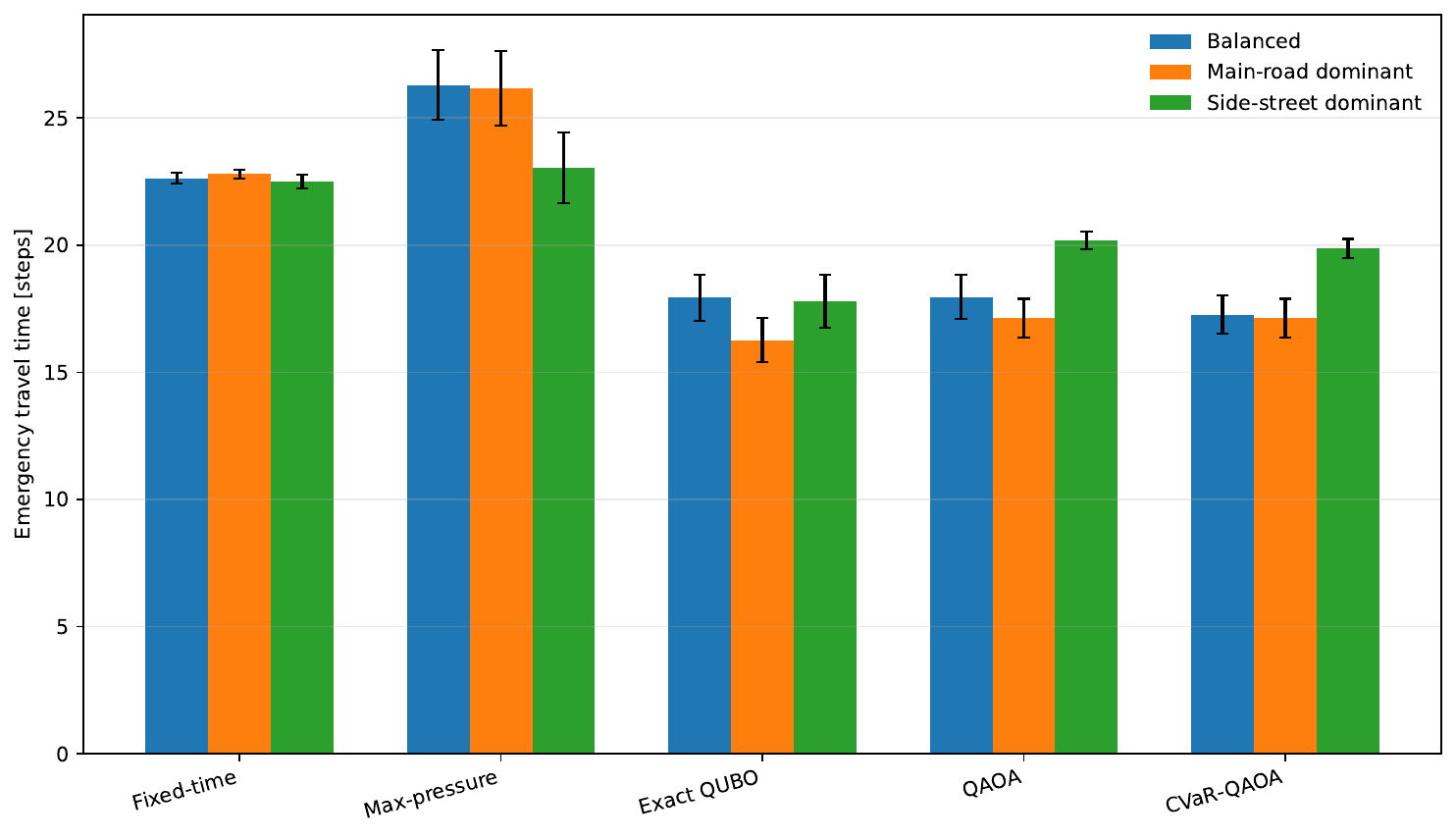}
    \caption{Emergency-vehicle travel time for each controller under the three traffic-demand regimes. Error bars show 95\% confidence intervals.}
    \label{fig:rq1_emergency_travel}
\end{figure}

Emergency travel times for successful runs are reported in
Fig.~\ref{fig:rq1_emergency_travel} and Tab.~\ref{tab:rq1_summary}.
Under balanced demand, the mean travel time decreased from 22.63 steps
for fixed-time control and 26.30 steps for max-pressure to 17.93, 17.97,
and 17.27 steps for Exact \gls{qubo}, \gls{qaoa}, and \gls{cvarqaoa}, respectively.
A similar pattern was observed under main-road-dominant demand, where
Exact \gls{qubo} achieved the lowest mean travel time of 16.27 steps, compared
with 17.13 steps for both \gls{qaoa} and \gls{cvarqaoa}.

Under side-street-dominant demand, Exact \gls{qubo} retained a mean emergency
travel time of 17.80 steps, while \gls{qaoa} and \gls{cvarqaoa} required 20.20 and
19.87 steps, respectively. Both nevertheless maintained 100\% completion,
unlike max-pressure, which completed only 63.3\% of the runs within the
simulation horizon.

These results indicate that explicitly incorporating emergency priority
into the finite-horizon optimization improves emergency progression and
provides consistent completion across the tested demand regimes.

\subsubsection{Ordinary-Traffic Performance}

\begin{figure}[t]
    \centering
    \includegraphics[width=\linewidth]{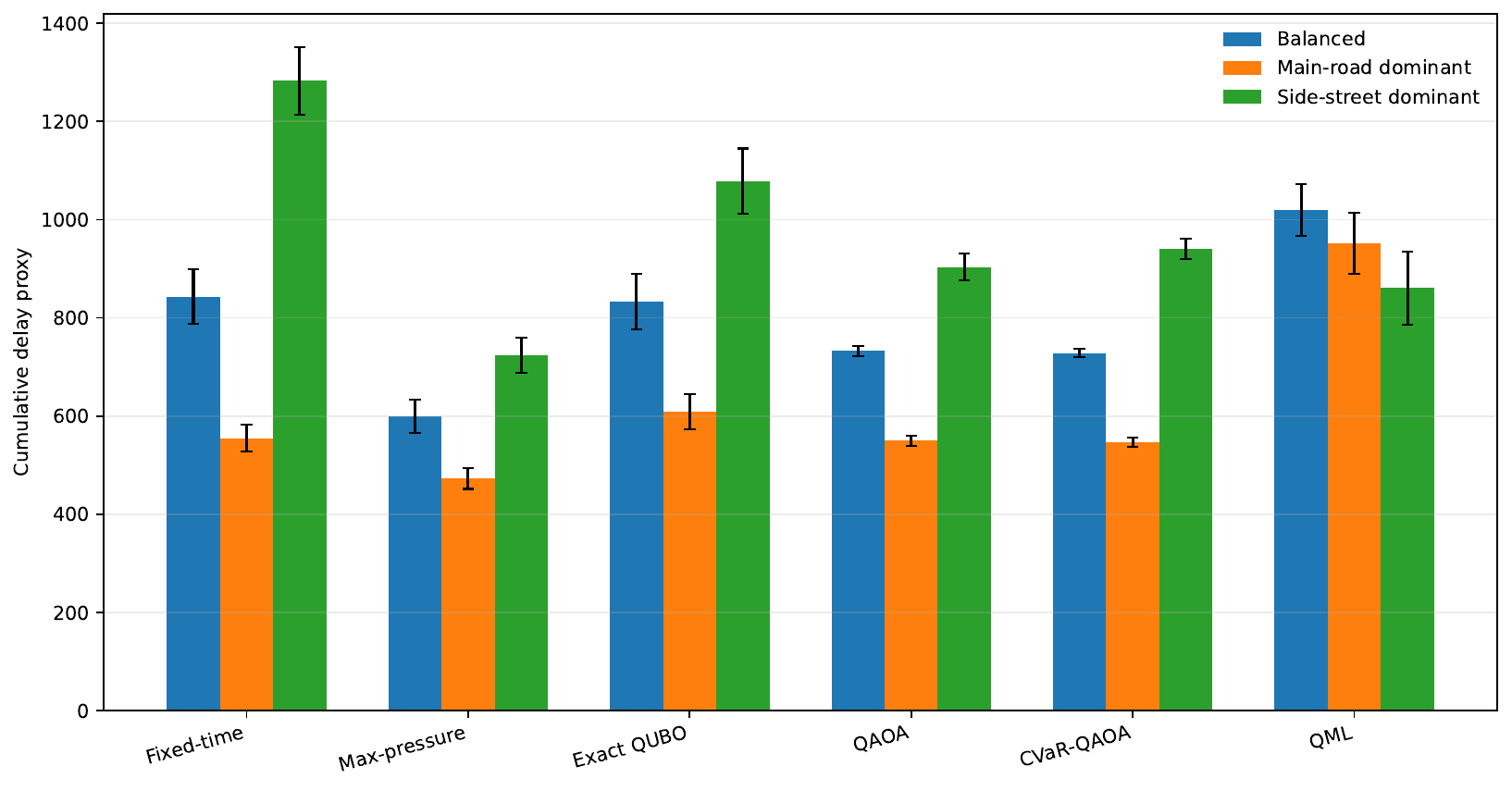}
    \caption{Cumulative traffic-delay proxy for each controller under the three traffic-demand regimes. Error bars show 95\% confidence intervals.}
    \label{fig:rq1_cumulative_delay}
\end{figure}

Ordinary-traffic performance showed a different ranking from emergency-vehicle performance. As shown in Fig.~\ref{fig:rq1_cumulative_delay} and Tab.~\ref{tab:rq1_summary}, max-pressure produced the lowest cumulative delay in all three demand regimes, with mean values of 599.55 under balanced demand, 473.15 under main-road-dominant demand, and 723.58 under side-street-dominant demand.

The optimization-based controllers generally reduced cumulative delay relative to fixed-time control, although they did not outperform max-pressure. Under balanced demand, cumulative delay decreased from 843.30 for fixed-time control to 832.95 for Exact \gls{qubo}, 732.83 for \gls{qaoa}, and 728.43 for \gls{cvarqaoa}. The largest improvement relative to fixed-time control occurred under side-street-dominant demand, where \gls{qaoa} and \gls{cvarqaoa} reduced the mean cumulative delay from 1282.97 to 903.31 and 940.17, respectively.

A similar pattern was observed for side-street queues. Max-pressure achieved the lowest mean side-street queue length under all three demand regimes, while \gls{qaoa} and \gls{cvarqaoa} generally remained below fixed-time control. These results indicate that the emergency-priority optimization improves ordinary-traffic performance relative to fixed-time control in several conditions, but the queue-responsive max-pressure controller remains the strongest method for minimizing ordinary congestion.

\subsubsection{Emergency--Congestion Trade-off}

The results reveal a clear trade-off between emergency progression and ordinary-traffic efficiency. Max-pressure consistently achieved the lowest cumulative delay and shortest side-street queues, but its emergency-vehicle performance was weaker, particularly under balanced and side-street-dominant demand. In contrast, Exact \gls{qubo}, \gls{qaoa}, and \gls{cvarqaoa} maintained 100\% emergency completion across all demand regimes and generally reduced emergency travel time, at the cost of higher ordinary-traffic delay than max-pressure.

Among the optimization-based controllers, \gls{qaoa} and \gls{cvarqaoa} provided the most balanced compromise. Under balanced demand, for example, \gls{cvarqaoa} reduced mean emergency travel time to 17.27 steps while keeping cumulative delay at 728.43, compared with 26.30 steps and 599.55 for max-pressure. Under side-street-dominant demand, \gls{qaoa} and \gls{cvarqaoa} preserved 100\% emergency completion while keeping cumulative delay substantially below fixed-time control. These results indicate that the optimization-based methods trade some ordinary-traffic efficiency for substantially more reliable emergency progression.

\subsubsection{\glsentryshort{qml} Performance}

The \gls{qml} policy was substantially less reliable than the optimization-based controllers. Its emergency-completion rate was 63.3\% under balanced demand, 13.3\% under main-road-dominant demand, and 40.0\% under side-street-dominant demand. The reported emergency travel times therefore reflect only the subset of runs in which the emergency vehicle successfully completed the corridor.

This conditional reporting introduces a survivorship effect. For example, the apparently competitive travel time under side-street-dominant demand is computed only from the 12 successful runs out of 30 and should not be interpreted as evidence of overall competitive emergency performance. The substantially lower completion rates indicate that the learned policy did not reliably reproduce the sequential behavior required for emergency progression.

The limited reliability is consistent with the moderate imitation performance obtained during training. Because the policy is trained offline to reproduce only the first action of the Exact \gls{qubo} controller, classification errors can accumulate over successive control steps and alter the subsequent traffic state. The present \gls{qml} configuration should therefore be regarded as a preliminary imitation-based controller rather than a competitive alternative to the optimization-based methods.

\subsubsection{Longer-Horizon Robustness}

The 72-step robustness experiment largely confirmed the controller ranking observed in the principal 36-step evaluation. Exact \gls{qubo}, \gls{qaoa}, \gls{cvarqaoa}, fixed-time, and max-pressure completed the emergency trip in all 10 runs under each demand regime. In contrast, the \gls{qml} policy remained less reliable, completing all 10 runs under balanced demand but only 6 of 10 under main-road-dominant demand and 8 of 10 under side-street-dominant demand.

Emergency travel times remained close to those observed in the shorter experiment. Under balanced demand, the mean travel times were 22.6 steps for fixed-time, 26.5 for max-pressure, 18.1 for Exact \gls{qubo}, 18.1 for \gls{qaoa}, and 17.4 for \gls{cvarqaoa}. Under main-road-dominant demand, the corresponding values were 22.8, 26.0, 16.2, 17.4, and 17.4 steps, while under side-street-dominant demand they were 22.4, 25.0, 18.1, 20.1, and 20.0 steps.

The ordinary-traffic trade-off also persisted over the longer horizon. Max-pressure again produced the lowest cumulative delay in all three demand regimes, whereas \gls{qaoa} and \gls{cvarqaoa} remained substantially below fixed-time control under balanced and side-street-dominant demand. These results indicate that the main conclusions of \glsentryshort{rq1} are not specific to the 36-step simulation horizon.

\begin{table}[t]
\centering
\caption{\glsentryshort{rq1} performance summary across the three traffic-demand regimes. Emergency travel time is reported only for runs in which the emergency vehicle completed the corridor.}
\label{tab:rq1_summary}
\resizebox{\linewidth}{!}{
\begin{tabular}{llrrrr}
\hline
Demand & Controller & Completion [\%] & Em. travel & Cum. delay & Side queue \\
\hline
Balanced
& Fixed-time   & 100.0 & 22.63 & 843.30 & 6.99 \\
& Max-pressure & 90.0  & 26.30 & 599.55 & 4.51 \\
& Exact \gls{qubo}   & 100.0 & 17.93 & 832.95 & 6.56 \\
& \gls{qaoa}         & 100.0 & 17.97 & 732.83 & 5.66 \\
& \gls{cvarqaoa}    & 100.0 & \textbf{17.27} & 728.43 & 5.54 \\
& \gls{qml}          & 63.3  & 23.84 & 1019.04 & 8.00 \\
\hline
Main-road dominant
& Fixed-time   & 100.0 & 22.80 & 555.13 & 4.04 \\
& Max-pressure & 96.7  & 26.17 & \textbf{473.15} & \textbf{3.08} \\
& Exact \gls{qubo}   & 100.0 & \textbf{16.27} & 609.80 & 4.68 \\
& \gls{qaoa}         & 100.0 & 17.13 & 550.05 & 4.25 \\
& \gls{cvarqaoa}    & 100.0 & 17.13 & 546.37 & 4.31 \\
& \gls{qml}          & 13.3  & 23.00 & 951.92 & 4.49 \\
\hline
Side-street dominant
& Fixed-time   & 100.0 & 22.50 & 1282.97 & 11.36 \\
& Max-pressure & 63.3  & 23.05 & \textbf{723.58} & \textbf{5.88} \\
& Exact \gls{qubo}   & 100.0 & \textbf{17.80} & 1078.54 & 9.20 \\
& \gls{qaoa}         & 100.0 & 20.20 & 903.31 & 7.45 \\
& \gls{cvarqaoa}    & 100.0 & 19.87 & 940.17 & 7.89 \\
& \gls{qml}          & 40.0  & 20.33 & 860.55 & 6.07 \\
\hline
\end{tabular}
}
\end{table}
\subsection{\glsentryshort{rq2}: Emergency-Priority Trade-off}

\subsubsection{Priority-Weight Sweep}
The emergency-priority sweep evaluates $\lambda_{\mathrm{em}}\in\{0,5,10,20,35,50,75\}$ for Exact \gls{qubo}, \gls{qaoa}, and \gls{cvarqaoa} under balanced demand. Each configuration was evaluated over the same set of 30 traffic seeds, allowing the effect of increasing emergency priority to be compared across the three optimization-based controllers. Tab.~\ref{tab:rq2_priority} summarizes the resulting emergency travel time, emergency waiting time, cumulative delay, and mean side-street queue size.

\subsubsection{Emergency Travel Time}

\begin{figure}[t]
    \centering
    \includegraphics[width=\linewidth]{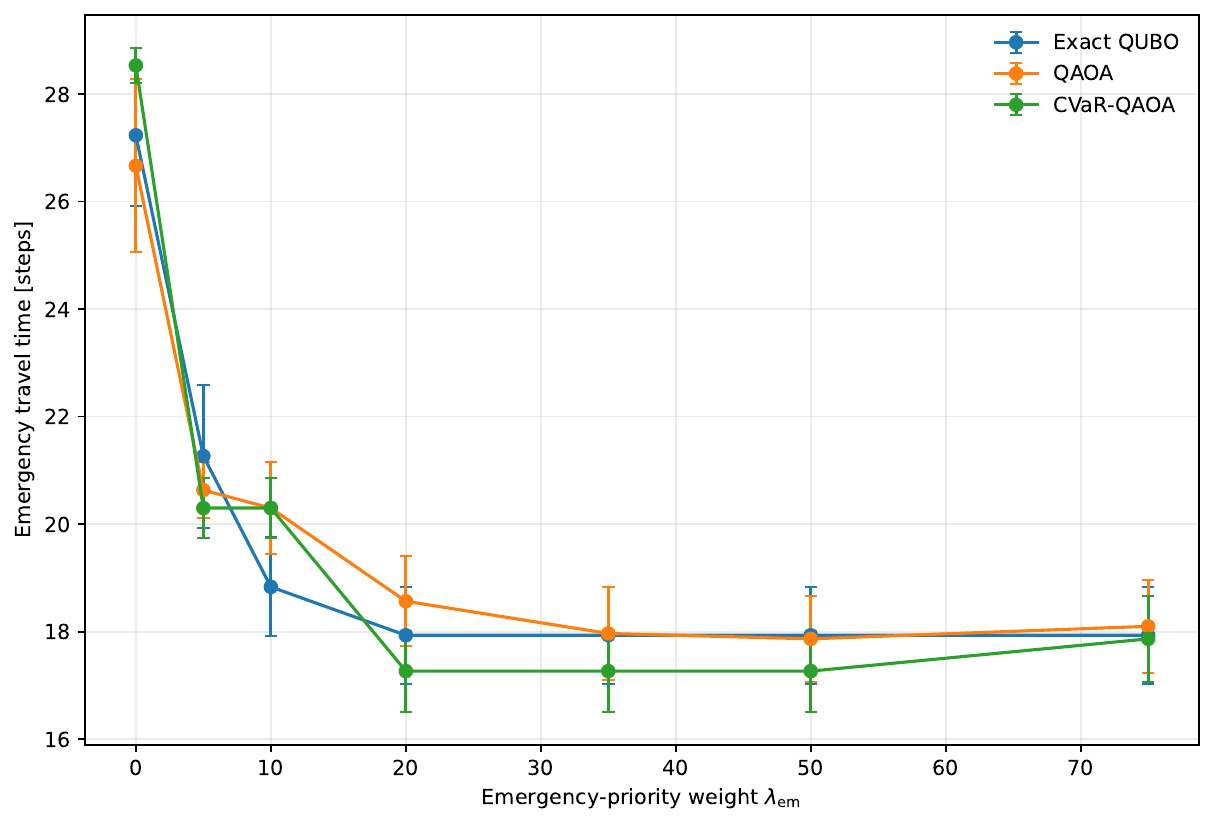}
    \caption{Emergency-vehicle travel time as a function of the emergency-priority weight $\lambda_{\mathrm{em}}$ under balanced demand. Error bars show 95\% confidence intervals.}
    \label{fig:rq2_emergency_travel}
\end{figure}

Emergency travel time decreased markedly as the emergency-priority weight increased from zero. As shown in Fig.~\ref{fig:rq2_emergency_travel} and Tab.~\ref{tab:rq2_priority}, Exact \gls{qubo} decreased from 27.23 steps at $\lambda_{\mathrm{em}}=0$ to 21.27 at $\lambda_{\mathrm{em}}=5$ and 18.83 at $\lambda_{\mathrm{em}}=10$, reaching 17.93 steps at $\lambda_{\mathrm{em}}=20$. \gls{qaoa} showed a similar reduction from 26.67 steps at $\lambda_{\mathrm{em}}=0$ to 18.57 at $\lambda_{\mathrm{em}}=20$, while \gls{cvarqaoa} decreased from 28.53 to 17.27 steps over the same range.

Beyond $\lambda_{\mathrm{em}}=20$, additional improvements were small. Exact \gls{qubo} remained at 17.93 steps for all larger tested weights, while \gls{cvarqaoa} remained at 17.27 steps through $\lambda_{\mathrm{em}}=50$ and increased slightly to 17.87 at $\lambda_{\mathrm{em}}=75$. \gls{qaoa} varied only modestly between 17.87 and 18.10 steps over the larger weights. The results therefore indicate that most of the emergency-travel-time benefit is achieved by approximately $\lambda_{\mathrm{em}}=20$, after which the response largely saturates.

\subsubsection{Emergency Waiting Time}

Emergency waiting time showed a stronger floor effect than travel time. As reported in Tab.~\ref{tab:rq2_priority}, \gls{qaoa} achieved a mean waiting time of 0.00 steps for every tested value of $\lambda_{\mathrm{em}}$. For \gls{cvarqaoa}, the mean waiting time decreased from 2.80 steps at $\lambda_{\mathrm{em}}=0$ to 0.00 steps at all nonzero tested weights. Exact \gls{qubo} decreased from 3.70 steps at $\lambda_{\mathrm{em}}=0$ to 0.77 at $\lambda_{\mathrm{em}}=5$ and 0.07 at $\lambda_{\mathrm{em}}\geq10$.

Because of the many identical zero or near-zero waiting times, this metric contains numerous ties and provides less discrimination among the controllers than emergency travel time. The priority-weight effect is therefore more clearly reflected by travel time and ordinary-traffic cost than by waiting time alone. 

\subsubsection{Ordinary-Traffic Cost}

\begin{figure}[t]
    \centering
    \includegraphics[width=\linewidth]{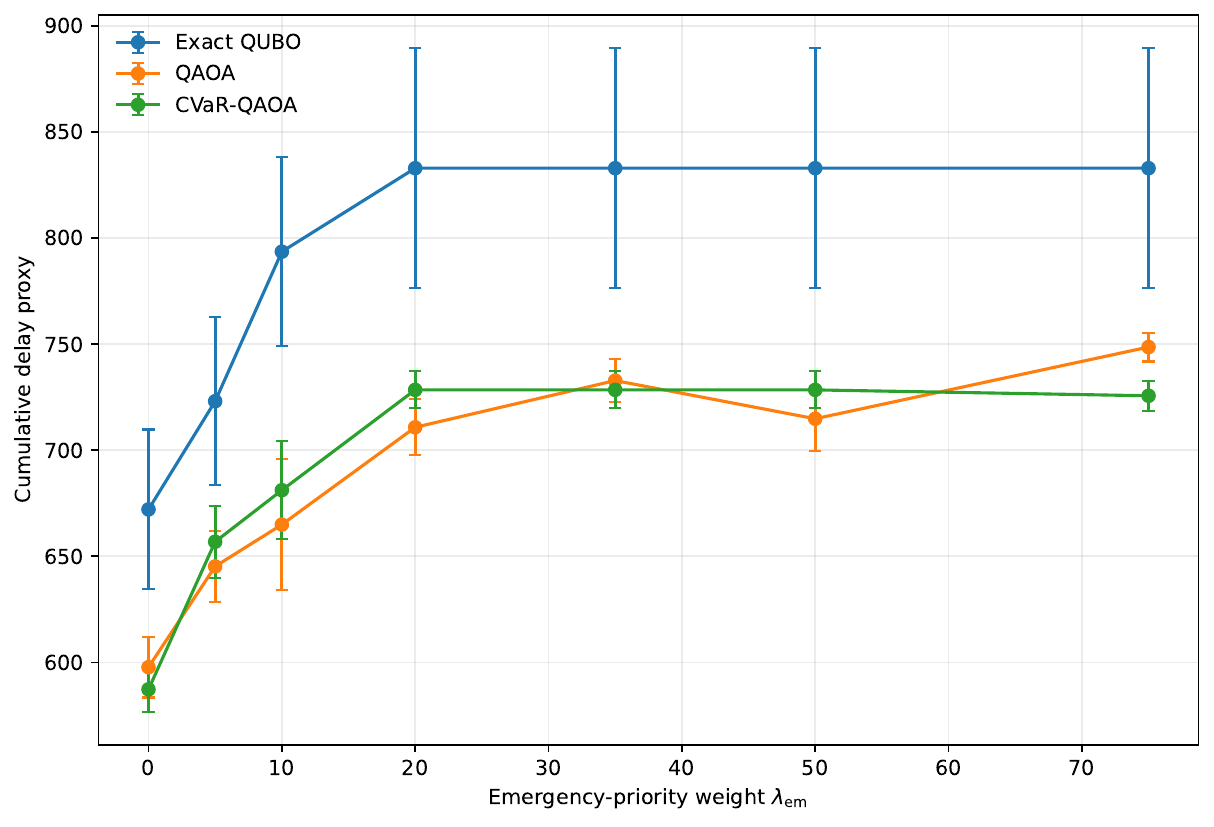}
    \caption{Cumulative traffic-delay proxy as a function of the emergency-priority weight $\lambda_{\mathrm{em}}$ under balanced demand. Error bars show 95\% confidence intervals.}
    \label{fig:rq2_delay}
\end{figure}

The reduction in emergency travel time was accompanied by higher ordinary-traffic cost. As shown in Fig.~\ref{fig:rq2_delay} and Tab.~\ref{tab:rq2_priority}, the mean cumulative delay for Exact \gls{qubo} increased from 672.08 at $\lambda_{\mathrm{em}}=0$ to 793.58 at $\lambda_{\mathrm{em}}=10$ and 832.95 at $\lambda_{\mathrm{em}}=20$, after which it remained unchanged for the larger tested weights. A similar, although less pronounced, increase was observed for the quantum controllers. For \gls{qaoa}, cumulative delay increased from 597.71 at $\lambda_{\mathrm{em}}=0$ to 710.81 at $\lambda_{\mathrm{em}}=20$, while \gls{cvarqaoa} increased from 587.28 to 728.43 over the same interval.

\begin{figure}[t]
    \centering
    \includegraphics[width=\linewidth]{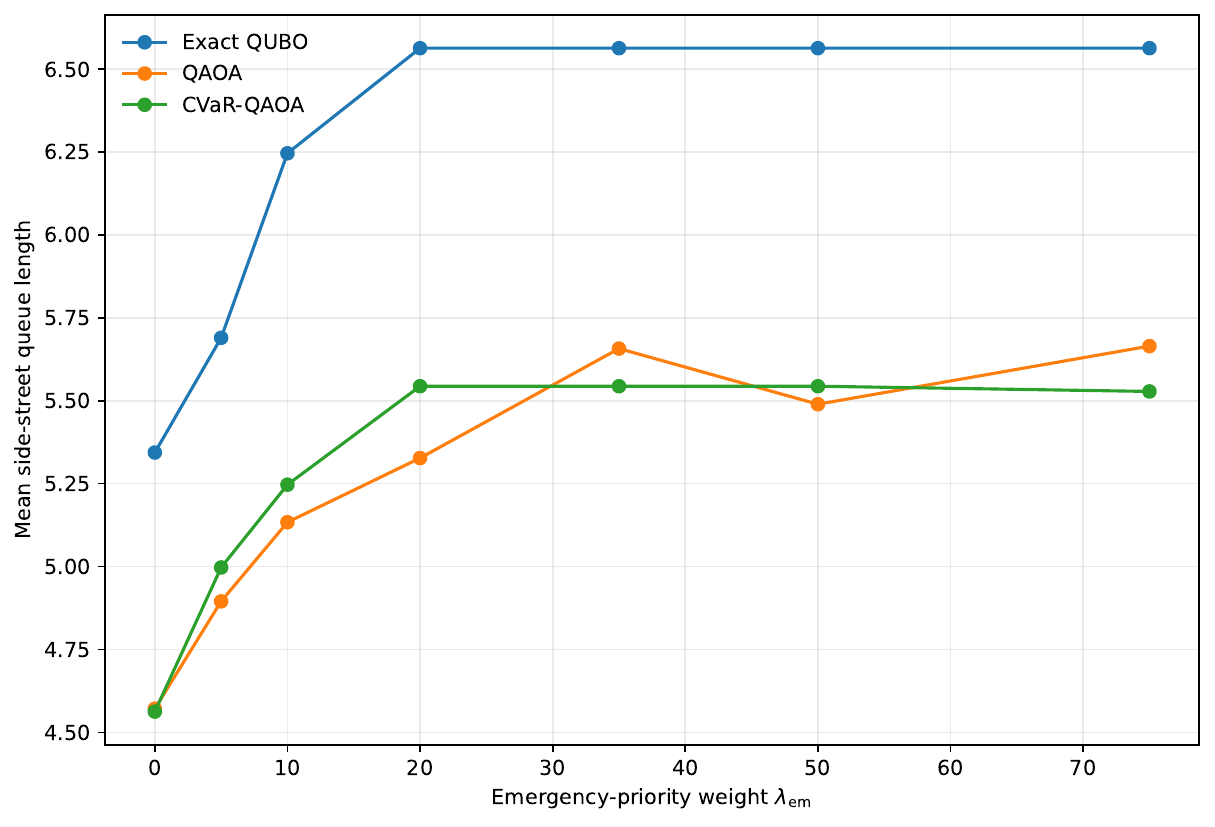}
    \caption{Mean side-street queue length as a function of the emergency-priority weight $\lambda_{\mathrm{em}}$ under balanced demand.}
    \label{fig:rq2_side_queue}
\end{figure}

The side-street queues exhibited the same overall trade-off, as shown in Fig.~\ref{fig:rq2_side_queue}. For Exact \gls{qubo}, the mean side-street queue increased from 5.34 at $\lambda_{\mathrm{em}}=0$ to 6.56 at $\lambda_{\mathrm{em}}=20$. For \gls{qaoa}, it increased from 4.57 to 5.33, while for \gls{cvarqaoa} it increased from 4.56 to 5.54. At larger priority weights, the quantum-controller values fluctuated within a relatively narrow range rather than increasing monotonically. These results show that stronger emergency priority shifts signal capacity toward the main corridor, reducing emergency travel time at the expense of increased ordinary-traffic congestion.

\begin{table}[t]
\centering
\caption{\glsentryshort{rq2} emergency-priority sweep under balanced demand. Values are means over 30 matched traffic seeds.}
\label{tab:rq2_priority}
\resizebox{\linewidth}{!}{
\begin{tabular}{rllrrr}
\hline
$\lambda_{\mathrm{em}}$ & Controller & Em. travel & Em. wait & Cum. delay & Side queue \\
\hline
0
& Exact \gls{qubo}  & 27.23 & 3.70 & 672.08 & 5.34 \\
& \gls{qaoa}        & 26.67 & 0.00 & 597.71 & 4.57 \\
& \gls{cvarqaoa}   & 28.53 & 2.80 & 587.28 & 4.56 \\
\hline
5
& Exact \gls{qubo}  & 21.27 & 0.77 & 723.10 & 5.69 \\
& \gls{qaoa}        & 20.63 & 0.00 & 645.22 & 4.90 \\
& \gls{cvarqaoa}   & 20.30 & 0.00 & 656.85 & 5.00 \\
\hline
10
& Exact \gls{qubo}  & 18.83 & 0.07 & 793.58 & 6.25 \\
& \gls{qaoa}        & 20.30 & 0.00 & 664.94 & 5.13 \\
& \gls{cvarqaoa}   & 20.30 & 0.00 & 681.14 & 5.25 \\
\hline
20
& Exact \gls{qubo}  & 17.93 & 0.07 & 832.95 & 6.56 \\
& \gls{qaoa}        & 18.57 & 0.00 & 710.81 & 5.33 \\
& \gls{cvarqaoa}   & \textbf{17.27} & 0.00 & 728.43 & 5.54 \\
\hline
35
& Exact \gls{qubo}  & 17.93 & 0.07 & 832.95 & 6.56 \\
& \gls{qaoa}        & 17.97 & 0.00 & 732.83 & 5.66 \\
& \gls{cvarqaoa}   & \textbf{17.27} & 0.00 & 728.43 & 5.54 \\
\hline
50
& Exact \gls{qubo}  & 17.93 & 0.07 & 832.95 & 6.56 \\
& \gls{qaoa}        & 17.87 & 0.00 & 714.83 & 5.49 \\
& \gls{cvarqaoa}   & \textbf{17.27} & 0.00 & 728.43 & 5.54 \\
\hline
75
& Exact \gls{qubo}  & 17.93 & 0.07 & 832.95 & 6.56 \\
& \gls{qaoa}        & 18.10 & 0.00 & 748.63 & 5.67 \\
& \gls{cvarqaoa}   & \textbf{17.87} & 0.00 & 725.66 & 5.53 \\
\hline
\end{tabular}
}
\end{table}

\subsubsection{Priority Saturation}

The results indicate a saturation region beginning at approximately $\lambda_{\mathrm{em}}=20$. By this point, Exact \gls{qubo}, \gls{qaoa}, and \gls{cvarqaoa} had already achieved most of their reduction in emergency travel time, while further increases in the priority weight produced only small changes. Exact \gls{qubo}, for example, remained at 17.93 steps for all tested values from $\lambda_{\mathrm{em}}=20$ to $75$, while \gls{cvarqaoa} remained at 17.27 steps between $\lambda_{\mathrm{em}}=20$ and $50$.

At the same time, the corresponding cumulative-delay and side-street-queue values remained above those obtained at low priority weights. The range
\begin{equation}
20 \lesssim \lambda_{\mathrm{em}} \lesssim 35
\end{equation}
therefore captures the region in which most of the emergency-progression benefit has already been obtained without requiring still larger emergency-priority coefficients. The default value $\lambda_{\mathrm{em}}=35$ used in the other experiments lies within this saturation region.

\subsection{\glsentryshort{rq3}: Quantum Approximation Quality}
\glsentryshort{rq3} evaluates how closely \gls{qaoa} and \gls{cvarqaoa} reproduce the exact finite-horizon \gls{qubo} solution and whether differences in \gls{qubo} quality translate into differences in closed-loop traffic performance. The analysis considers the optimality gap, exact-solution hit rate, probability assigned to globally optimal states, and agreement on the first action that is actually applied by the receding-horizon controller.
\subsubsection{Optimality Gap}

\begin{figure}[t]
    \centering
    \includegraphics[width=\linewidth]{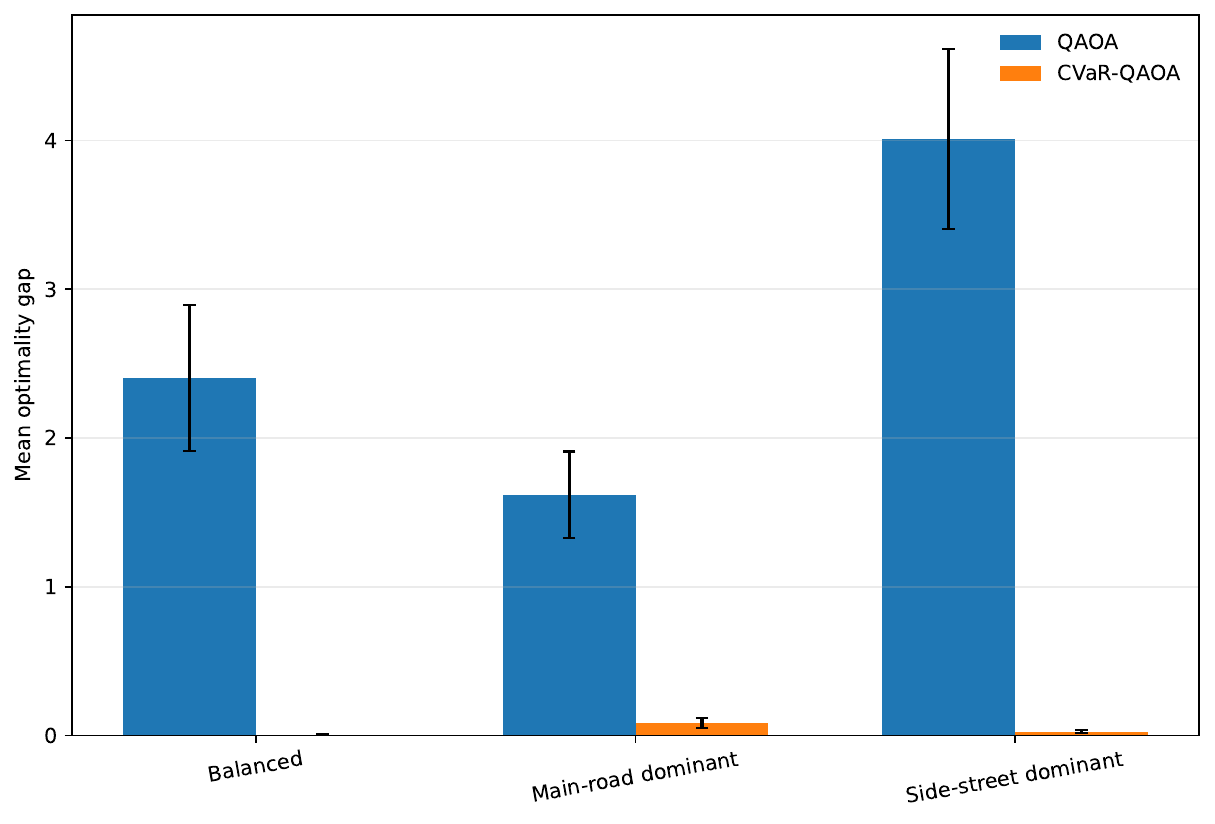}
    \caption{Mean \gls{qubo} optimality gap of \gls{qaoa} and \gls{cvarqaoa} relative to the exact \gls{qubo} solution across the three traffic-demand regimes. Error bars show 95\% confidence intervals.}
    \label{fig:rq3_optimality_gap}
\end{figure}

The mean \gls{qubo} optimality gap differed substantially between the two quantum optimizers. As shown in Fig.~\ref{fig:rq3_optimality_gap} and Tab.~\ref{tab:rq3_quality}, standard \gls{qaoa} produced mean gaps of 2.401 under balanced demand, 1.619 under main-road-dominant demand, and 4.008 under side-street-dominant demand. In contrast, \gls{cvarqaoa} reduced the corresponding mean gaps to 0.006, 0.085, and 0.026.

The largest difference occurred under side-street-dominant demand, where the mean gap decreased from 4.008 for \gls{qaoa} to 0.026 for \gls{cvarqaoa}. Across all three demand regimes, \gls{cvarqaoa} therefore produced solutions much closer in energy to the exact \gls{qubo} optimum than standard \gls{qaoa}.

\subsubsection{Exact-Solution Hit Rate}

\begin{figure}[t]
    \centering
    \includegraphics[width=\linewidth]{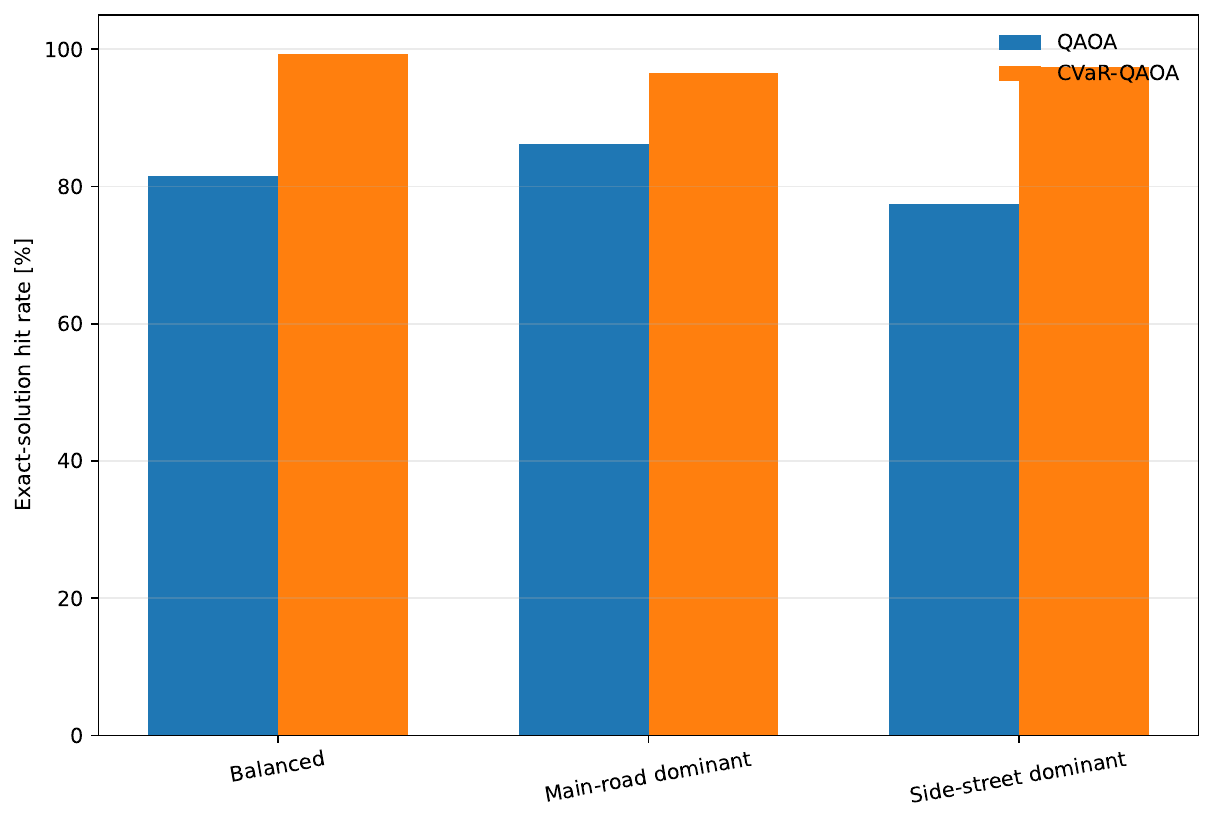}
    \caption{Exact-\gls{qubo} solution hit rate of \gls{qaoa} and \gls{cvarqaoa} across the three traffic-demand regimes.}
    \label{fig:rq3_exact_hit}
\end{figure}

The exact-solution hit rate shows how often the quantum optimizer selected a bitstring with the same \gls{qubo} energy as the global optimum. As shown in Fig.~\ref{fig:rq3_exact_hit} and Tab.~\ref{tab:rq3_quality}, \gls{qaoa} achieved hit rates of 81.57\% under balanced demand, 86.20\% under main-road-dominant demand, and 77.41\% under side-street-dominant demand. The corresponding values for \gls{cvarqaoa} were substantially higher at 99.35\%, 96.57\%, and 97.41\%, respectively.

The difference was especially pronounced under balanced and side-street-dominant demand, where \gls{cvarqaoa} matched the exact \gls{qubo} optimum in nearly all optimization steps. These results are consistent with the optimality-gap analysis and show that \gls{cvarqaoa} reproduced the exact finite-horizon \gls{qubo} solution more reliably than standard \gls{qaoa}.

\subsubsection{Optimal-State Probability}

The probability assigned to globally optimal states was also consistently higher for \gls{cvarqaoa}. As reported in Tab.~\ref{tab:rq3_quality}, the mean optimal-state probability under balanced demand increased from 5.25\% for \gls{qaoa} to 11.90\% for \gls{cvarqaoa}. Under main-road-dominant demand, the corresponding values were 3.98\% and 11.16\%, while under side-street-dominant demand they were 4.94\% and 14.75\%.

Thus, although the probability mass assigned to globally optimal solutions remained relatively small in absolute terms, \gls{cvarqaoa} concentrated substantially more probability on optimal states than standard \gls{qaoa} across all three demand regimes. This is consistent with both its smaller optimality gaps and its higher exact-solution hit rates.

\subsubsection{First-Action Agreement}

Because the controller operates in a receding-horizon manner, agreement on the first action is particularly important: only the first decision of each optimized horizon is actually applied before the \gls{qubo} is rebuilt. Despite the larger differences observed in full-schedule \gls{qubo} quality, both quantum controllers showed high first-action agreement with the exact solution.

As reported in Tab.~\ref{tab:rq3_quality}, \gls{qaoa} achieved mean first-action agreement rates of 98.15\% under balanced demand, 98.52\% under main-road-dominant demand, and 97.62\% under side-street-dominant demand. The corresponding values for \gls{cvarqaoa} were 100.00\%, 100.00\%, and 99.78\%, respectively. These results indicate that even when the complete bitstring selected by \gls{qaoa} was suboptimal, its immediately executed action frequently matched that of the exact \gls{qubo} solution.
\begin{figure}[t]
    \centering
    \includegraphics[width=\linewidth]{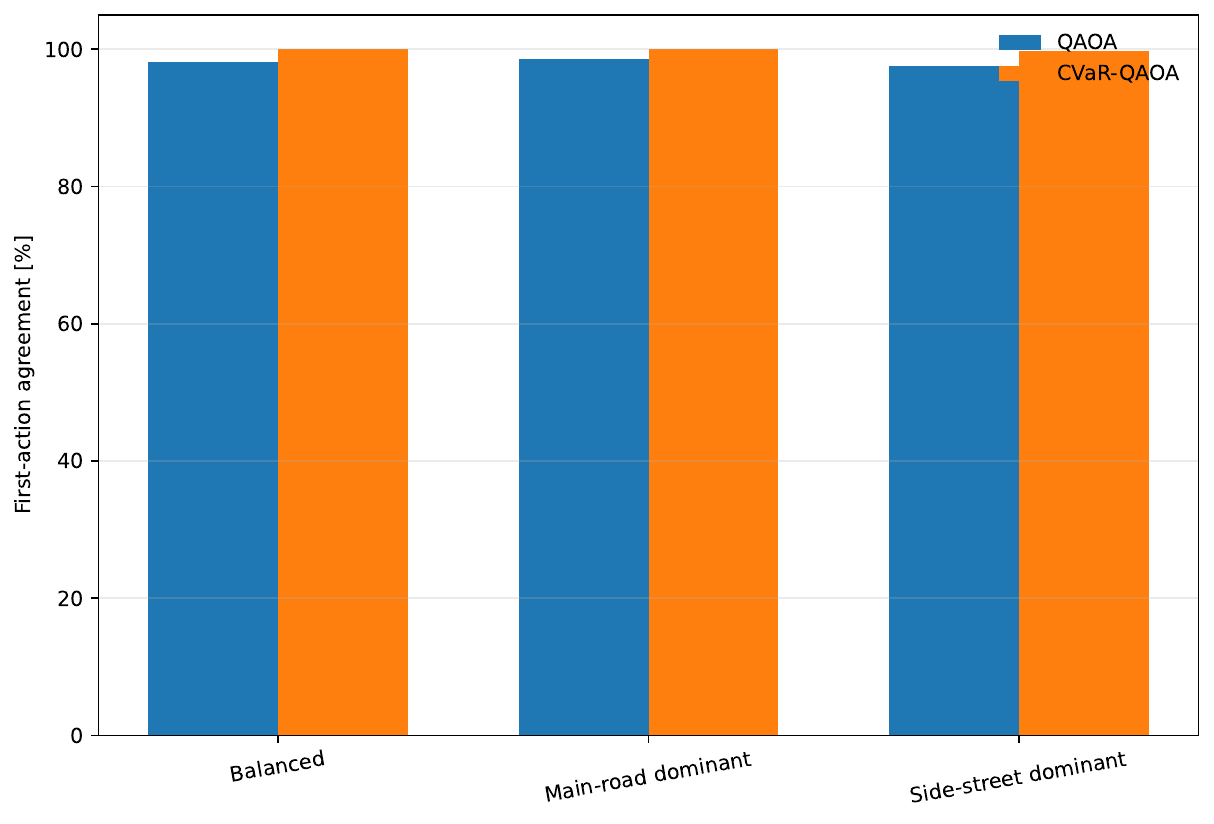}
    \caption{Agreement between the first actions selected by \gls{qaoa} and \gls{cvarqaoa} and the exact \gls{qubo} solution across the three traffic-demand regimes.}
    \label{fig:rq3_first_action}
\end{figure}

\subsubsection{\glsentryshort{qubo} Quality vs. Traffic Performance}

The substantial improvement in \gls{qubo}-solution quality obtained with \gls{cvarqaoa} did not translate into proportionally large changes in closed-loop traffic performance. For example, under balanced demand the mean optimality gap decreased from 2.401 for \gls{qaoa} to 0.006 for \gls{cvarqaoa}, while the corresponding emergency travel times changed only from 17.97 to 17.27 steps and cumulative delay from 732.83 to 728.43. Under main-road-dominant demand, both quantum controllers produced the same mean emergency travel time of 17.13 steps despite the exact-solution hit rate increasing from 86.20\% for \gls{qaoa} to 96.57\% for \gls{cvarqaoa}.

This difference arises because the \gls{qubo} energy is a finite-horizon surrogate rather than the realized cumulative traffic objective. Moreover, only the first optimized action is applied at each simulation step, and the safety supervisor may further constrain the requested phase. Consequently, different full-horizon bitstrings can produce the same executed control action, and a lower \gls{qubo} energy does not necessarily imply a proportional reduction in realized traffic delay. The \gls{cvarqaoa} results therefore demonstrate substantially closer reproduction of the exact \gls{qubo} solution, while the resulting closed-loop traffic differences relative to standard \gls{qaoa} are considerably smaller.
\begin{table}[t]
\centering
\caption{\glsentryshort{rq3} \gls{qubo}-solution quality of \gls{qaoa} and \gls{cvarqaoa}. Hit rate, optimal-state probability, and first-action agreement are reported in percent.}
\label{tab:rq3_quality}
\resizebox{\linewidth}{!}{
\begin{tabular}{llrrrr}
\hline
Demand & Controller & Mean gap & Exact hit [\%] &
$P_{\mathrm{opt}}$ [\%] & First action [\%] \\
\hline
Balanced
& \gls{qaoa}      & 2.401 & 81.57 & 5.25 & 98.15 \\
& \gls{cvarqaoa} & \textbf{0.006} & \textbf{99.35} & \textbf{11.90} & \textbf{100.00} \\
\hline
Main-road dominant
& \gls{qaoa}      & 1.619 & 86.20 & 3.98 & 98.52 \\
& \gls{cvarqaoa} & \textbf{0.085} & \textbf{96.57} & \textbf{11.16} & \textbf{100.00} \\
\hline
Side-street dominant
& \gls{qaoa}      & 4.008 & 77.41 & 4.94 & 97.62 \\
& \gls{cvarqaoa} & \textbf{0.026} & \textbf{97.41} & \textbf{14.75} & \textbf{99.78} \\
\hline
\end{tabular}
}
\end{table}

\section{Discussion}

The results show a consistent trade-off between emergency-vehicle progression and ordinary-traffic efficiency. In \glsentryshort{rq1}, Exact \gls{qubo}, \gls{qaoa}, and \gls{cvarqaoa} maintained 100\% emergency completion across all tested demand regimes and generally reduced emergency travel time relative to fixed-time and max-pressure control. Max-pressure, however, produced the lowest cumulative delay and shortest side-street queues, indicating that explicitly prioritizing emergency progression incurs an ordinary-traffic cost. The longer-horizon experiment preserved the same qualitative pattern, suggesting that these observations are not specific to the 36-step simulation horizon. The \gls{qml} policy was substantially less reliable, particularly under main-road- and side-street-dominant demand, which indicates that the present offline imitation approach is not yet robust enough for sequential deployment.

The \glsentryshort{rq2} priority sweep clarifies how this trade-off is controlled by the emergency-priority coefficient. Increasing $\lambda_{\mathrm{em}}$ strongly reduced emergency travel time between $\lambda_{\mathrm{em}}=0$ and approximately $\lambda_{\mathrm{em}}=20$, while cumulative delay and side-street queues increased over the same range. Beyond approximately $\lambda_{\mathrm{em}}=20$--$35$, additional emergency-travel improvements were small, indicating diminishing returns from further increasing the priority weight. The default value $\lambda_{\mathrm{em}}=35$ therefore lies in a region where most of the observed emergency-progress benefit has already been obtained, although the resulting choice remains application dependent because different deployments may assign different costs to emergency delay and ordinary congestion.

The strongest quantum-specific result arises from \glsentryshort{rq3}. Across all three demand regimes, \gls{cvarqaoa} produced substantially smaller \gls{qubo} optimality gaps, higher exact-solution hit rates, and greater probability on globally optimal states than standard \gls{qaoa}. Nevertheless, these large differences in optimization quality resulted in much smaller differences in realized traffic performance. This is expected because the optimized \gls{qubo} objective is only a finite-horizon surrogate for the closed-loop traffic process. Only the first action of each optimized schedule is executed before replanning, and the safety supervisor can further constrain that action. Consequently, different full-horizon solutions can generate the same executed decision, and a lower \gls{qubo} energy does not necessarily imply proportionally lower realized traffic delay.

These results should therefore be interpreted as evidence about algorithmic behavior rather than quantum computational advantage. The experiments use a small three-intersection corridor, synthetic traffic demand, statevector simulation, and a nine-qubit \gls{qubo}. Execution on physical quantum processors introduces additional considerations,
including gate and measurement imperfections and device-dependent behavior, which motivate explicit hardware characterization and benchmarking \cite{lewandowska2025benchmarking,bilek2026experimental}.  Within this scope, however, the results show that \gls{cvarqaoa} is a substantially more accurate solver for the tested \gls{qubo} formulation than standard \gls{qaoa}, while also illustrating that improvements in optimization fidelity must be evaluated jointly with their effect on the closed-loop control objective.

\section{Reproducibility}

The complete implementation, experiment scripts, configuration files, raw and processed result data, and analysis outputs used in this study are made publicly available through a Zenodo replication package \cite{illesova2026hybrid_traffic}. The archive includes the datasets underlying \glsentryshort{rq1}, \glsentryshort{rq2}, and \glsentryshort{rq3}, including the longer-horizon robustness runs and the step-level quantum-optimization diagnostics required to reproduce the reported figures and tables. 

\section{Conclusion}

This work investigated coordinated emergency-priority traffic control using a finite-horizon \gls{qubo} formulation together with Exact \gls{qubo}, \gls{qaoa}, \gls{cvarqaoa}, and \gls{qml} controllers. For \glsentryshort{rq1}, the optimization-based controllers consistently achieved reliable emergency progression across the tested demand regimes, while max-pressure remained more effective at minimizing ordinary-traffic congestion. For \glsentryshort{rq2}, increasing the emergency-priority weight reduced emergency travel time but increased cumulative delay and side-street queues, with most of the observed benefit reached by approximately $\lambda_{\mathrm{em}}=20$--$35$. For \glsentryshort{rq3}, \gls{cvarqaoa} reproduced the exact \gls{qubo} solution substantially more accurately than standard \gls{qaoa}, although these improvements in optimization quality translated into smaller differences in closed-loop traffic performance. The \gls{qml} policy was less reliable and requires further development. Overall, the results demonstrate the feasibility of the proposed optimization framework while highlighting the importance of evaluating both surrogate \gls{qubo} quality and realized traffic performance. Future work should consider larger networks, richer traffic models, and execution on noisy quantum hardware.
\printglossaries
\bibliographystyle{splncs04}
\bibliography{references}

\end{document}